ARTICLE

# Quantitative Modelling of Solvation in 2D Materials via Alchemical Decoupling and Potential Energy Surfaces

Lindsey N. Ioos,[a] Duncan A. Peterson,[a] and Scott C. Warren[a]*

Understanding why 2D materials remain suspended in solutions is essential to understanding the mechanisms of aggregation and exfoliation. These dynamics have been difficult to model because of the long timescales and large activation energies involved in these simulations. Here we describe an improved strategy for modelling the solvation and stability of 2D materials that uses the Gibbs free energy of solvation to accurately calibrate two disconnected potential energy surfaces for the fully desolvated and fully solvated 2D materials. By joining these surfaces using solvation energy, we obtain predictions for 2D material stability in various solvents that are qualitatively different than prior simulations and that have excellent agreement with experimental results. Our approach predicts a much higher barrier to full desolvation of the material, suggesting that materials that appear to be aggregated in fact still retain monolayers of solvent between adjacent layers of 2D materials. This work provides much needed understanding on the interactions between the solvent and 2D material that can guide future experimental work.

## Introduction

Investigations into 2D materials have risen in recent decades from increased interest in their unique characteristics. Since the discovery of graphene in 2004[1] a number of new 2D materials have been synthesized, each exhibiting different properties than their bulk counterparts.[2-8] Many methods exist for creating 2D materials, but liquid phase exfoliation (LPE) remains one of the most popular for its affordability and scalability.[9-15] LPE involves breaking apart weak van der Waals forces holding together the bulk material and creating a suspension of the isolated layers in a solvent. A solvent shell forms around the monolayers, maintaining their stability once exfoliated.[16] The choice of solvent is an important part of the synthesis, since it determines the effectiveness of the exfoliation and dispersion.[17-19] However, the interactions between the 2D material and the solvent are complex and efforts at quantifying these interactions have been limited.[14,18,20,21] Further, understanding the dynamics that facilitate 2D material stability in solution has been challenging due to the long timescales and the large activation energies involved.

Experimental studies have systematically characterized suspensions of 2D materials, hoping to identify improved solvents and to recognize trends that might give insight into the microscopic mechanisms of exfoliation and aggregation.[22,23] A study by Coleman et al. measured UV-Vis absorbance to examine the concentration of 2D materials in solvents with different surface tensions and found that some of these solvents were able to disperse more material than others.[14] No statistically significant trends between dispersion quality and surface tension were found. Other research efforts have explored zeta potentials and electrostatic effects,[24,25] surface energy components,[21] and other solubility parameters such as the Hildebrand and Hansen parameters.[26,27] Hansen parameters provide a detailed way to quantify solvent-solute interactions but many studies[14,20] show that Hansen parameters have limited predictive value for solvent screening for 2D materials. The challenges in these experiments, due to the difficulty of isolating individual variables, have implications for finding and predicting new solvents for 2D materials. These issues have led some researchers to pursue a computational approach.

Molecular dynamics simulations have been shown effective in modelling the forces that govern these intermolecular interactions in 2D materials, such as graphene.[28] Simulations allow for the isolation of properties such as solvent diffusion[29] and binding and exfoliation energies.[30] A recent study by Blankschtein and coworkers used potential mean force (PMF) calculations to create potential energy surfaces for sheets of $MoS_2$ in different solvents.[31] The energy was calculated between the two sheets upon approach, mapping the stability of the layered structure upon exfoliation and predicting resulting configurations. An initial model for exfoliation was proposed in which solvents with higher energy barriers created more stable dispersions, such as N-methyl pyrrolidine (NMP). This method has since been used by others to investigate surfactant interactions with $MoS_2$ dispersions,[32] and has been applied to phosphorene,[34] indicating its applicability to other 2D materials.[35]

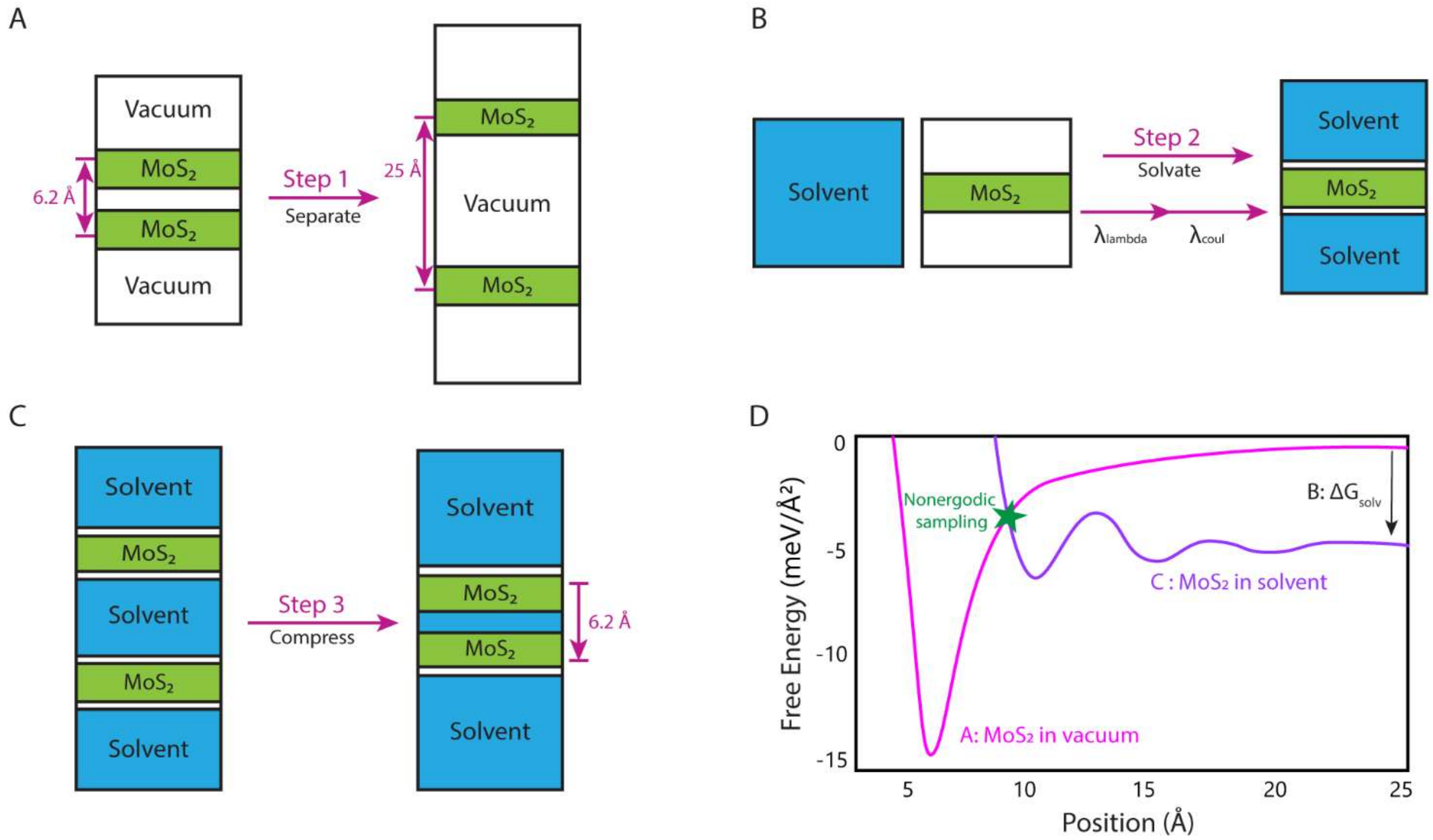

**Fig. 1:** Three step method for calculating PMF curves. (A) Step one involves calculating the energy of separating two sheets in a vacuum. (B) Once separated, the energy to solvate the sheets are calculated. (C) The sheets are then brought together while immersed in solvent. (D) This process calibrates the broken energy surfaces to create a full PMF curve.

Recent experiments for controlled aggregation studies identified results inconsistent with predictions from the PMF simulations.[35] When solvent was evaporated from suspensions of 2D materials, the resulting materials had a single monolayer of solvent trapped between 2D materials. However, the potential energy surfaces from prior work predicted that solvent bilayers were often more stable.[31] Peterson et al. discovered that this discrepancy resulted from the 2D sheets bending into unrealistic geometries in the transition state.[35] When Peterson et al. restricted the bending of the $MoS_2$, they found poor sampling during the highest energy transition state. Thus, they proposed using the solvation energy ($\Delta G_{solv}$) to reconnect the surfaces and demonstrated the efficacy of this new method for one of the solvents explored in the previous work, NMP.

The current work expands the range of solvents discussed by Peterson et al. to those most commonly used in $MoS_2$ exfoliation (DMF, DMSO, IPA, and $H_2O$) and compares them to the original work by Blankschtein and coworkers. Upon running the modified simulations, it was found that the potential energy surfaces are qualitatively different. We found larger barriers that prevent the full desolvation of $MoS_2$ in most solvents, which leads us to conclude that structures with a single layer of solvent between layers of material are more favorable, and more likely to be seen experimentally. These results align with the reported d-spacings of solvent intercalated materials in several studies.[36,37] Our analysis shows how van der Waals interactions can be further characterized using computational methods and give insight into the exfoliation of 2D materials.

## Methods

Building on the umbrella sampling approach introduced by Blankschtein and coworkers, we observed that transition-states remained challenging to sample because $MoS_2$ sheets could bend into unphysical configurations. Increasing the force constants during sampling reduced these unphysical configurations but adequate sampling of the full reaction coordinate, especially in the transition states, was still not possible. To address these sampling challenges, we adopted an alternative approach for estimating the potential energy surfaces. In this section, we describe the overall methodology together with the specific methods used for PMF and solvation energy calculations.

To create a calibrated potential energy surface (Fig. 1D), we perform two disconnected PMF simulations—one in vacuum (Fig. 1A) and one in solvent (Fig. 1C). We then relate the energies in vacuum to those in solvent using a solvation energy calculation (Fig. 1B). This calibrates the PMF curves to accurately predict potential energy across most of the reaction coordinate, despite poor sampling in the transition state when the final solvent layer is expelled from between $MoS_2$ sheets.

**PMF Simulation Details**

The PMF simulations calculate the energy between two sheets of $MoS_2$ as they approach each other. All simulations were performed under NPT conditions. The PMF calculations were performed with two AB stacked $MoS_2$ sheets oriented parallel to each other within a 7 x 7 x 9 nm solvent filled simulation box. Each monolayer was hexagonally shaped to minimize edge effects and had a length of 4.1 nm edge to edge (Fig. 2A). The sheets start 2.5 nm apart, a large enough distance where there are limited interactions. The simulations begin with annealing the sheets at 400 $^oC$ for 5 ns while using a spring to hold the sheets in place. Throughout the PMF simulation, the two sheets were held to a parallel orientation using a spring restraint of 120,000 kcal/mol/$Å^2$ on a per-atom basis for each layer.[35] This was applied to the x, y, and z axes for the bottom layer, and the x and y axes for the top layer, allowing it to move up and down the z axis freely while maintaining parallelism.

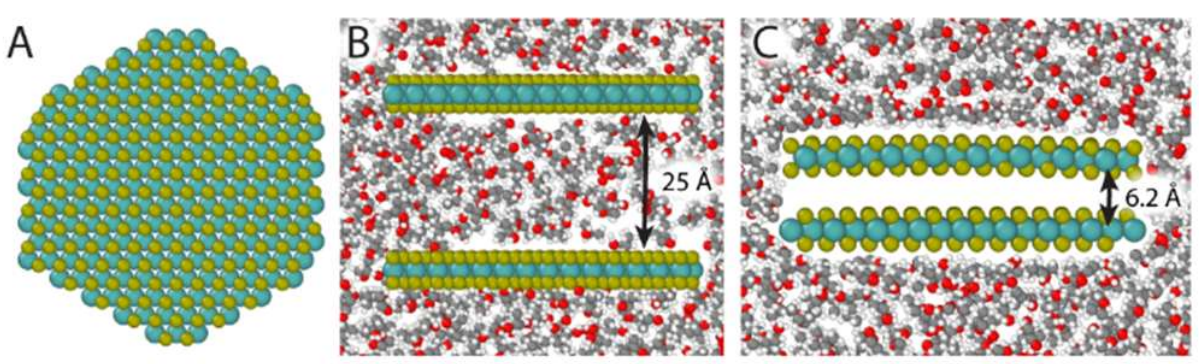

**Fig. 2:** PMF simulation details. (A) The hexagonal $MoS_2$ sheet used for simulations, 4.1 nm edge to edge length. (B) Sheets separated at maximum distance used in the simulation (25 Å) with solvent between. (C) Minimum distance between sheets (6.2Å), resulting in solvent ejected from the structure.

The COLVARS module[38] in LAMMPS was used to sample the reaction coordinate as the $MoS_2$ sheets were pulled together or apart (Fig. 2B). The simulations sampled center-to-center separations from 0.5 nm to 2.5 nm. A smaller force constant of 300 kcal/mol/$Å^2$ was used for larger separation (2.5 to 1.5 nm), while a larger force constant of 3000 kcal/mol/$Å^2$ was used for smaller separations (1.5 to 0.5 nm). Sampling was performed every 0.1 Å for large separations and every 0.025 Å for small separations. Each position was sampled for 1.5 million timesteps, while trajectory information was recorded every 5 timesteps. Each simulation runs for 300 million timesteps. The SHAKE algorithm in LAMMPS was used to constrain hydrogen bonds, allowing the simulation to run at larger timesteps while still modelling the effects of the hydrogen atoms.[39] A weighted histogram analysis, BayesWHAM[40], was used to create the potential energy surface. The first 30% of timesteps were excluded from each position, minimizing hysteresis between simulations performed in either direction. Binning was performed along the z-axis with a bin width of 0.001 Å. Calculations converged to $10^{-15}$.

For a quantitative analysis of the number of molecules present at each configuration throughout the simulation, an areal density plot was created for each solvent. OVITO[41] was used to visualize the simulation cells, and the number of molecules between the two sheets was counted at each 0.1 Å change in the simulation. Analyzing the changes in the number of molecules at specific points allows us to characterize the structures formed at different stages.

**Solvation Energy of 2D Materials**

The solvation energy of a sheet of $MoS_2$ in solvent was calculated using an alchemical decoupling method developed by Khanna et al.[42] described in Fig. 3. With these simulations, the coupling parameter, λ, was varied to analyze the effects of the Leonard-Jones (LJ) and coulombic interactions between the solvent and solute.

Decoupling calculations begin with the LJ and coulombic forces from the simulation being isolated and turned off (λ=0). The LJ forces are then incrementally turned back on, causing the repulsion in the system to grow. Once fully turned on (λ=1), the coulombic forces are also incrementally turned on, while the LJ forces remain on throughout. The Gibbs free energy change is calculated along these pathways for each set of force fields and combined to calculate the full free energy of solvation for the system This relationship is described through equations 1 and 2 below, where *i* indicates LJ or coulombic force fields.

$$\Delta \boldsymbol{G}_{solv} = \Delta \boldsymbol{G}_{LJ} + \Delta \boldsymbol{G}_{coul} \quad (1)$$

$$\Delta \boldsymbol{G}^{i}_{solv} = \int_{\lambda_i=0}^{\lambda_i=1} \left\langle \frac{dU}{d\lambda} \right\rangle d\lambda_i \quad (2)$$

The solvation energy simulations used a rectangular 3.8 nm by 4.0 nm sheet of $MoS_2$, containing 546 atoms. The sheet spanned the width and depth of the simulation cell, and periodic boundary conditions were used to connect the edges of the sheet. Therefore, no interactions between the solvent and the sheet edges occured. The x, y, and z dimensions of pressure were decoupled to allow the simulation cell to expand and compress along the z axis as needed, while maintaining the dimension of the sheet along x and y. These dimensions were controlled independently using the $P_{xx}$, $P_{yy}$, and $P_{zz}$ components of the stressor tensor. All simulations were performed under NPT conditions. Each system was equilibrated for 10 ns and ran for 8 ns for each LJ or coulomb calculation. The LJ integration was done using a three-part Gauss-Legendre quadrature, 7 points from 0 to a, twenty points from a to b, and 13 points from b to 1. The parameters a and b were adjusted for each simulation but typically ranged between 0.15-0.22. The coulombic integration used an 8-point quadrature from 0 to 1.

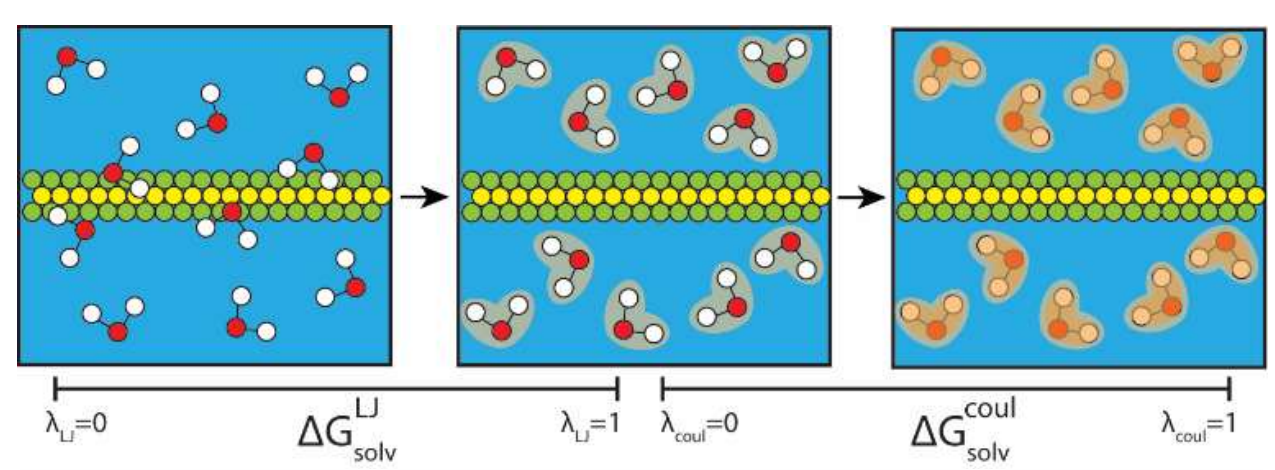

**Fig. 3**: The decoupling approach to calculating solvation energy. Beginning with all interactions turned off, LJ interactions are incrementally turned on. Once LJ interactions are fully turned on, the coulombic interactions are also incrementally turned on. The change in energy is calculated at each point, giving the total solvation energy.

## Results and discussion

**Free Energy of Solvation**

To determine the solvation energy of $MoS_2$ in various solvents, we performed an alchemical decoupling in which interactions between the solvent and $MoS_2$ were turned on in two steps. Initially, with the solvent-$MoS_2$ interactions turned off entirely, the molecules and $MoS_2$ physically overlap. In step 1, as the LJ interactions are turned on, the repulsive interactions between molecules and $MoS_2$ become strong enough that they no longer overlap. This transition is seen in Fig. 4A, in which the simulation energy rose as repulsive interactions grew, then suddenly dropped when the molecules no longer overlap. We tuned the quadrature sampling interval to ensure that this transition was well sampled. DMF, DMSO, and IPA were well sampled regardless of small changes to the quadrature (± 0.5%), but the abruptness of the transition for $H_2O$ led to increased difficulties in sampling and a larger uncertainty in the total solvation energy (± 10%). This could be due to the small size of the water molecules, or the extensive hydrogen bonding that is difficult to model.

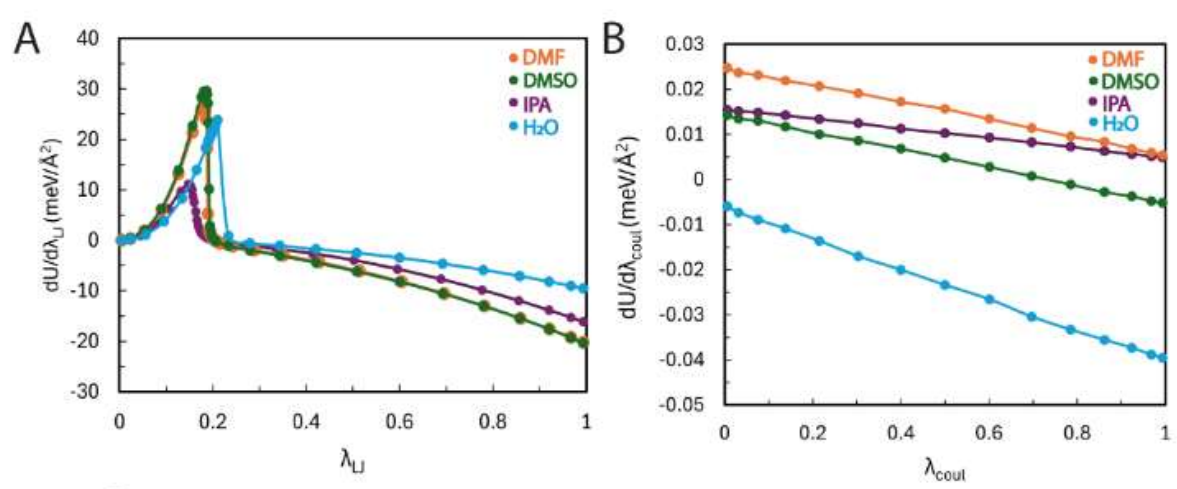

C

| Solvent | DMF | DMSO | IPA | $H_2O$ |
|---|---|---|---|---|
| $\Delta G_{solv}$ (meV/A$^2$) | -6.61 | -6.49 | -5.67 | -1.86 |

**Fig. 4**: Results of the alchemical decoupling method for calculating solvation energy for simulations of DMF, DMSO, IPA, and $H_2O$. (A) The LJ component is isolated and slowly turned on. (B) The coulombic component is then also turned on. (C) The total Gibbs free energy of solvation calculated.

The results for step 2 (turning on the coulombic interaction) show that the magnitude of the coulombic components are 10-100 times smaller than the LJ components. Fig. 4B shows that favorable interactions increase more for water as the coulombic component is turned on. It is important to note that the individual LJ and coulombic components are not state functions, so the differences between molecules have limited meaning. Only the total of the LJ and coulombic parts–the solvation energy (Fig. 4C)–has an interpretable meaning.

The total solvation energy calculated is presented in Fig. 4C and shows DMF with the largest free energy of solvation at -6.61 meV/Å$^2$, closely followed by DMSO, then IPA, and $H_2O$ with the smallest value of -1.86 meV/Å$^2$. These results show good agreement with experimental observations of solvent dispersions, where DMF ~ DMSO > IPA >> $H_2O$.[14] Typically, choosing solvents is based on the strength of the polarity of the material, with polar materials solvating best in polar solvents and vice versa. With 2D materials, it has been shown that solvents of intermediate polarity tend to be the best solvents.[14,15] These solvents are characterized by their modest dipoles and weaker hydrogen bonding, such as DMF and DMSO. For $MoS_2$ specifically, intermediate solvents containing nitrogen (NMP, DMF, DMSO) have been found to be good solvents due to the interactions between the nitrogen and sulfur vacancies on the surface of the material.

**Determination of Solvent Configurations**

The results of the PMF study are shown in Fig. 5. Both the PMF for separating two sheets in vacuum and the PMF for moving two sheets together in solvent are shown. The two curves are related by the solvation energy, described above. The areal density plot is also given. All of these curves show that the fully desolvated $MoS_2$ is the most stable configuration at 6.2 Å. This matches experimental data for restacked $MoS_2$, which has a center-to-center spacing of 6.15 Å.[43] As the desolvated $MoS_2$ sheets were separated in vacuum, the energy rose and asymptotically arrived at a stable value when the sheets were fully separated, defined as 0 meV/Å$^2$. As the solvated $MoS_2$ sheets were separated in solvent, the energy oscillated as the distance between them increased, resulting in qualitatively different curves for each simulation.

To interpret the oscillations in the solvated potential energy surfaces for each solvent, we first visualized the solvent configuration at different points along the potential energy surface (Fig. 5). For the simulation in DMF (Fig. 5A), the first minima is at 10 Å. Visualization shows that the DMF molecules form a single monolayer with the molecules lying horizontally (h) and an areal density of 0.038 molecules/Å$^2$. The second minimum occurs at 12.3 Å. Visualization of the simulation shows that the molecules still lie in a single layer but are now oriented vertically (v) with an areal density of 0.055 molecules/Å$^2$. A third minima occurs at a separation of 16.1 Å, and visualization shows that the molecules are now arranged in a somewhat disordered bilayer structure with an areal density of 0.088 molecules/Å$^2$. The approximate doubling (130% increase) of the areal density is consistent with the bilayer containing about twice as many molecules as the horizontal monolayer.

Proposing a general model to explain all solvent interactions with the material is challenging. However, there are trends that emerge across all four data sets. In particular, barrier height correlates with the number of molecules lost over a given increment of distance: a rapid decrease in the number of molecules is associated with higher energy barriers, whereas a more gradual decrease corresponds to lower barriers. For water, the final energy barrier is low, corresponding with a facile loss of water molecules for full desolvation. For the other solvents, the final barrier is high enough that the simulations cannot overcome it, indicating that full desolvation is very unlikely. Water is known to be a poor solvent for $MoS_2$, as indicated by its relatively low solvation energy. As the water system is the only simulation to demonstrate full desolvation, this result suggests that solvation energy may influence the barrier height separating the monolayer and the full loss of solvent.

The presence of energy minima at monolayer and bilayer spacings indicates that solvent structure plays a role in stabilizing particular configurations. Differences in low energy configurations (e.g., as a vertical monolayer vs. horizontal monolayer) are subtle, given the similarity in their energies that are seen. One factor that could explain some of these differences is the relatively strong dipole moment of DMSO or DMF, which favors vertical stacking by allowing the dipoles on adjacent molecules to point in opposite directions.[44] Regions of each simulation in fact showed this preference, potentially offering a low energy configuration. Evidence for a vertical monolayer can be found in the literature and seems to depend on the solvent-solvent forces present in the system. Pyridine is a common example of a vertical monolayer, with a reported d spacing of 12.3 Å in $MnPS_3$.[45] It interacts strongly with the surrounding solvent, and a vertical configuration maximizes the stabilizing effect of these interactions. Both DMF and DMSO offer a stable vertical monolayer, suggesting that solvent-solvent interactions play a larger role in stabilizing the configuration than expected.

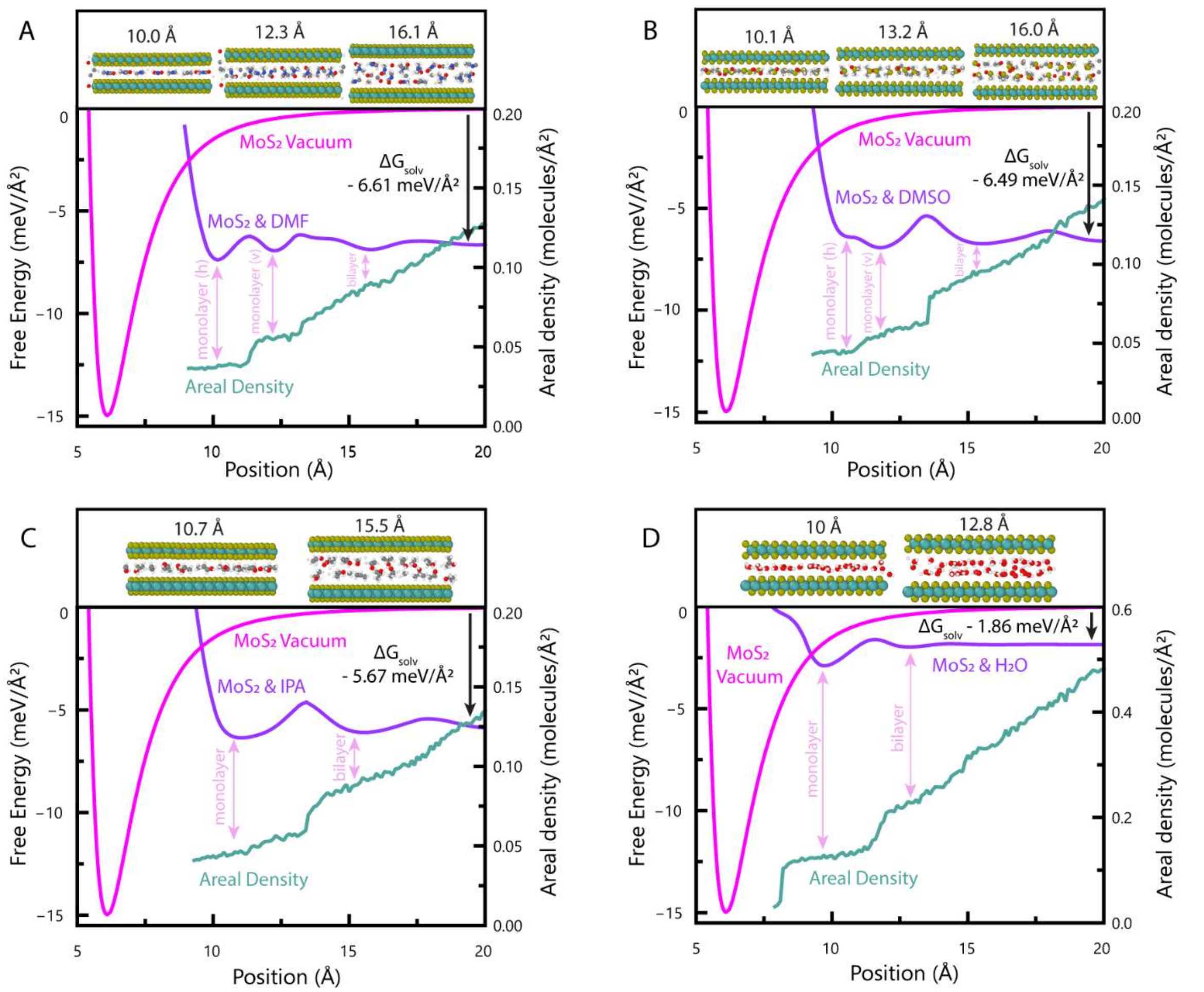

**Fig. 5:** The changes in the number of molecules between the two sheets of $MoS_2$ throughout the simulation are shown in (A) DMF, (B) DMSO, (C) IPA, and (D) $H_2O$. The stable structures, visualized as plateaus in the areal density curve and the minima from the potential energy surfaces, are labelled and a snapshot of the simulation solvent orientation is pictured.

**Potential Energy Surfaces (PMF Curves)**

The potential energy surfaces created with these simulations are compared to the previous work by Blankschtein and coworkers as seen in Fig. 6. For separations between sheets greater than 9 Å, there are qualitative differences between the two simulations. The energy values in our simulations are shifted lower due to the solvation energy calibration, and the absolute energy minima for our PMF curves are slightly shifted.

Where the two potential energy surfaces cross reveals the highest transition state across the reaction coordinate. Some caution is needed in interpreting the potential energy surface in this region, however, since the degree of mixing of between the two potential energy surfaces is not calculated; the true barrier height is lower than the crossing point of these surfaces. However, our predicted barrier for this transition state is consistently larger than that of the previous work. The lower barrier from the prior work indicates that full desolvation is likely for each solvent. Our model's higher barrier implies a more challenging transition to full desolvation, and thus the structure is more likely to remain as a monolayer in solvent instead of fully desolvating. This observation could explain why some solvents are better dispersants for $MoS_2$ than others: the energy barrier is too high for full desolvation to easily occur. This aligns with experimental observations of monolayers of $MoS_2$ dispersed in some solvents for extended amounts of time, while aggregating more quickly in others.

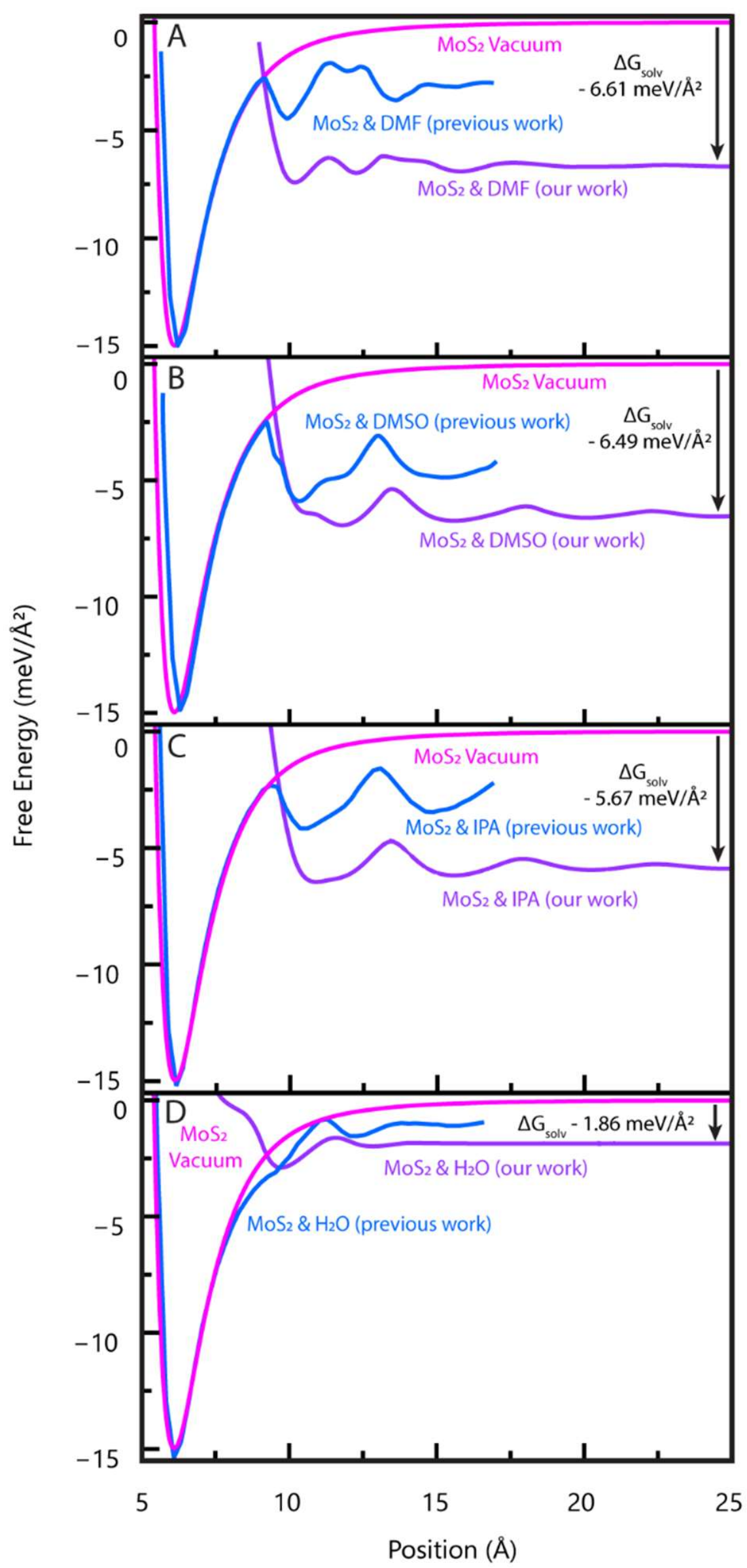

**Fig. 6:** PMF data for (A) DMF, (B) DMSO, (C) IPA, and (D) $H_2O$ compared to the previously collected data for each solvent from Blankschtein and coworkers. The solvation energy used to calibrate the system is given for each solvent.

For $MoS_2$ in DMF (Fig. 6A), the previous work predicts stable structures around 10 and 13.7 Å, with a less stable configuration around 11.9 Å. Our model shows the same minimum at 10 Å, but another metastable structure around 12.3 Å and a less stable structure at 16.1 Å. This agrees with experimental data of DMF monolayers occurring at 9.9 Å.[36] For $H_2O$ (Fig. 6D) the previous work shows a large minimum at 11.8 Å, and a small plateau at 9.6 Å. Our work has a small minimum at 13 Å and a more significant minimum at 9.5 Å. The difference in results occurs around the same transition state where sheet bending occurs. Experimental data shows monolayer $H_2O$ intercalation in $MoS_2$ expands the interlayer to 9.1 Å, matching with the prediction from our data of a more stable structure around 9.5 Å, instead of the 11.8 Å previously reported.[37] This indicates that our calculations are likely more realistic simulations than the previous work.

In Fig. 6B, showing $MoS_2$ in DMSO, there is a shifted stable configuration at 16 Å and stable structures at both 13.2 and 10.1 Å. Interestingly, the previous work predicts the monolayer configuration at lower interlayer distances to be more stable, the opposite of which is predicted in our simulations. The previous work also shows another configuration at 9.8 Å not shown in our data. Prior sampling protocols allowed the sheets to bend, resulting in the calculated data not occurring at the reported positionings. As these inconsistencies occur around the desolvation transition states, they can be attributed to the sheet bending that resulted from the previously conducted study. The graph for $MoS_2$ in IPA (Fig. 6C) is quite similar to that of the previous work, featuring two minima at 10.7 and 15.5 Å. The minima for IPA seem to match between our work and the previous work, but the barrier heights are different. This comparison shows that the two simulations predict different metastable configurations. The previous work has a large barrier between the minima at 15.5 Å and 10.7 Å, and a much smaller barrier between 10.7 Å and the desolvated state at 6.2 Å. We predict a much higher barrier between the minima at 10.7 Å and full desolvation implying a more metastable point at 10.7 Å than at 15.5 Å.

## Conclusion

We demonstrate that the solvation energy of 2D materials can be calculated in a variety of solvents using alchemical decoupling. The solvation energy values calculated were used to calibrate potential energy surfaces, providing an alternative method for understanding the changes in the solvation shell during the exfoliation of 2D materials. Our findings show that the molecular structure greatly influence the barrier heights. Therefore, examining the PMF curves allowed us to gain a mechanistic understanding of the features that allow 2D materials to remain dispersed in solvents. The PMF curves show a large barrier between a single layer of solvent and a fully desolvated 2D material, making it unlikely that a material will become fully desolvated and will instead stay suspended. The exception to this is water, which has a low barrier and indeed is known as a poor solvent for $MoS_2$. These results offer a significantly different model than offered in previous work.

Our approach of determining the solvation energy of 2D materials through alchemical decoupling may improve the ability to rationally develop solvents for processing 2D materials. Extensive prior work has sought to rationalize solvation characteristics of 2D materials using models such as Hansen and Hildebrand solubility parameters. Now having quantitative values for solvation energy should allow the validity of such models to be assessed using quantitative metrics.

While our work only reports simulations of $MoS_2$, the methods can be used to quantify other 2D materials as well, such as phosphorene[14] and graphene.[9] These models show the importance of the final monolayer of solvent in keeping the sheets separated, justifying why the material stays solvated and exfoliates. A better understanding of the mechanism behind aggregation and exfoliation will allow us to build more complex 2D materials with advanced functionality.

## Author contributions

Investigation: L.I.; Simulation development: D.P., S.W.; all authors contributed to writing and editing the manuscript.

## Conflicts of interest

There are no conflicts to declare.

## Data availability

The data that supports the findings in this work, including the molecular dynamics simulations input files, analysis scripts, and raw output data, are available from the author upon reasonable request.

## Acknowledgements

This research was funded in part by a Summer Undergraduate Research Fellowship from the Office for Undergraduate Research at the University of North Carolina at Chapel Hill.